\documentclass[aps, prd, 10pt, twocolumn, superscriptaddress,noshowpacs, preprintnumbers, longbibliography,nofootinbib,bibnotes,hyperref,floatfix]{revtex4-2}
\usepackage[dvipsnames]{xcolor}
\usepackage[colorlinks=true,breaklinks=true]{hyperref}
\hypersetup{allcolors=[rgb]{0.0 0.0 0.70},linkcolor=blue}
\usepackage{orcidlink}
\usepackage{microtype}
\usepackage{float}
\usepackage{amsmath}
\usepackage{amsfonts}
\usepackage{amssymb}
\usepackage{mathtools}
\usepackage{multirow}
\usepackage{bm}
\usepackage[utf8]{inputenc}
\usepackage{graphicx}
\usepackage{dcolumn}
\usepackage{bm}
\usepackage{comment}

\usepackage{multirow}
\usepackage{booktabs}
\usepackage[normalem]{ulem}

\begin{document}
 \title{Quasi-steady flavor configuration of multi-energy neutrino ensembles}
\author{Manuel Goimil-Garc\'ia \orcidlink{0009-0001-0518-9274}}
 \affiliation{Niels Bohr International Academy \& DARK, Niels Bohr Institute,\\University of Copenhagen, Blegdamsvej 17, 2100 Copenhagen, Denmark}

\author{Irene Tamborra \orcidlink{0000-0001-7449-104X}}
\affiliation{Niels Bohr International Academy \& DARK, Niels Bohr Institute,\\University of Copenhagen, Blegdamsvej 17, 2100 Copenhagen, Denmark}

\date{\today}

\begin{abstract}
Neutrino flavor conversion profoundly impacts the explosion mechanism and multi-messenger emission of core-collapse supernovae. Yet, state-of-the-art hydrodynamic simulations of neutrino-dense astrophysical environments cannot account for neutrino quantum kinetics, necessitating subgrid schemes to model the impact of neutrino self-interaction on the quasi-steady flavor configuration. We present semi-analytic approximations for the outcomes of both slow and fast flavor conversions in quasi-homogeneous systems with periodic boundary conditions. Independent of the mass ordering, our ansatz demonstrates excellent agreement with multi-angle and multi-energy solutions of the neutrino kinetic equations across a wide range of representative (anti)neutrino distributions.
\end{abstract}

\maketitle

\section{Introduction}

Core-collapse supernovae and neutron-star merger remnants are  the most abundant neutrino sources besides the early universe~\cite{Janka:2025tvf,Burrows:2020qrp,Tamborra:2024fcd,Vitagliano:2019yzm,Raffelt:2025wty}. In such dense environments, neutrino refraction on electrons and  other neutrinos dramatically affects flavor conversion~\cite{Duan:2010bg,Mirizzi:2015eza,Tamborra:2020cul,Johns:2025mlm,Volpe:2023met}, with implications on the supernova explosion mechanism, the properties of compact remnants, and the nucleosynthesis of heavy elements~\cite{Ehring:2023lcd,Ehring:2023abs,Nagakura:2023mhr,Wang:2025nii,Wu:2017drk,Just:2022flt,George:2020veu,Lund:2025jjo,Qiu:2025kgy,Gogilashvili:2026epg,Gogilashvili:2026kef,Akaho:2026kff}.

The core of compact sources is characterized by extremely large neutrino and baryon densities. Here, flavor evolution is expected to be driven by collisional and fast instabilities~\cite{Sawyer:2005jk,Chakraborty:2016yeg,Johns:2021qby}. In particular, fast flavor conversion can occur when the electron-heavy  neutrino lepton number (ELN-XLN, where X stands for  muon or tau)  distribution has crossings~\cite{Izaguirre:2016gsx,Morinaga:2021vmc,Padilla-Gay:2021haz,Fiorillo:2023hlk,Fiorillo:2024dik}. This type of flavor transformation is called ``fast'' because its characteristic scale is linear in neutrino density.

Crossings between the energy spectra of different (anti)neutrino species can trigger ``slow" flavor conversion, depending on both the self-interaction strength and the frequency of vacuum oscillations. This phenomenon may occur at early post-bounce times in core-collapse supernovae, when ELN-XLN crossings are not present~\cite{Shalgar:2024gjt,Shalgar:2025oht,Fiorillo:2025gkw}, as well as  at large distances from the core of compact sources~\cite{Duan:2006an,Duan:2007bt,Fogli:2007bk,Fogli:2008pt}. {The latter possibility has been widely investigated within the bulb model, where (anti)neutrinos are steadily emitted from a spherical surface~\cite{Hannestad:2006nj,Duan:2006an,Duan:2007bt,Fogli:2007bk,Fogli:2008pt,Dasgupta:2009mg}. It was found that, in inverted ordering (IO), slow conversion swaps electron and heavy-lepton flavor (anti)neutrinos within the constraints of lepton number conservation, whereas collective effects in  normal ordering (NO) are negligible--see Refs.~\cite{Duan:2010bg,Mirizzi:2015eza} for dedicated reviews. On the contrary, neutrino ensembles in  homogeneous media~\cite{Raffelt:2007yz} or in boxes with periodic boundary conditions~\cite{Padilla-Gay:2025tko} can undergo slow instabilities in either mass ordering: the outcome of flavor conversion in these works (relying on single-energy ensembles)
depends on the neutrino-to-antineutrino ratio.} 
 In addition, the frequency of vacuum oscillations is expected to perturb the oscillation pattern of fast flavor transitions or effectively induce an ELN-XLN crossing for flavor configurations on the verge of becoming unstable~\cite{DedinNeto:2023ykt,Shalgar:2020xns,Fiorillo:2024uki}. 

To date, the neutrino kinetic equations cannot be solved in their full dimensionality because of the large gradients characterizing the quantities entering such equations. Global solutions of  flavor evolution focus on small spatial domains to track the development of slow and fast conversions and rely on  symmetry assumptions~\cite{Wu:2021uvt,Richers:2021nbx,Richers:2021xtf,Zaizen:2021wwl,Xiong:2024tac,Shalgar:2022lvv,Shalgar:2022rjj,Cornelius:2023eop,Shalgar:2024gjt,Cornelius:2024zsb,Shalgar:2025oht,Nagakura:2022kic,Nagakura:2022qko}. However, in the light of the potential impact of flavor conversion on the source  physics, nucleosynthesis, and multi-messenger observables, it is crucial to develop simplified approaches that would allow one to embed neutrino flavor conversion physics in hydrodynamic simulations. 

The first attempt in this direction was developed in  Ref.~\cite{Padilla-Gay:2021haz}, where  it was shown that the kinetic equations are  analogous to those of a gyroscopic pendulum for a homogeneous and azimuthally symmetric  system; hence, the overall amount of flavor conversion can be obtained relying on the real part of the eigenfrequency computed from the linear stability analysis  without solving the equations of motion.
Subgrid schemes could also serve such purpose for systems with more degrees of freedom, see e.g.~Refs.~\cite{Johns:2024dbe,Nagakura:2022xwe,Johns:2025yxa}. 
For example, building on the quasi-steady state flavor configurations coming from state-of-the-art numerical solutions of the neutrino kinetic equations, flavor equipartition  within the constraints of the lepton number and energy conservation was considered in Refs.~\cite{Ehring:2023lcd,Ehring:2023abs,Just:2022flt} to investigate the impact of flavor conversion on the physics of supernovae and neutron-star merger remnants. 

Alternative approaches model fast flavor conversion by means of an effective collision term in the Boltzmann equation, which drives the classical neutrino distribution towards an asymptotic state inferred from local simulations of neutrino flavor evolution~\cite{Liu:2025tnf,Zaizen:2022cik,Zaizen:2023ihz,Xiong:2021dex,Xiong:2024pue,Nagakura:2023jfi}. In Ref.~\cite{Goimil-Garcia:2025ozm}, we suggested a general approach to forecast the quasi-steady state flavor configuration of fast-unstable ensembles which can be easily implemented in such relaxation-time schemes, and highlighted the role of neutrino propagation in spreading flavor waves across spatial regions.
In this paper, we build on our previous findings to account for energy-dependent effects induced by the interplay between neutrino self-interaction and vacuum mixing, and consider both fast- and slow-unstable systems. 

This paper is organized as follows. The neutrino equations of motion and the system setup are introduced in Sec.~\ref{sec:eom}. We present our method to forecast the quasi-steady state of systems  with angular and/or energy crossings in Sec.~\ref{sec:ansatz}. In Sec.~\ref{sec:variations}, we explore the dependence of the quasi-steady state on the shape of the energy spectra and the mass ordering. We discuss our findings and conclude in Sec.~\ref{sec:conclusions}. 

\section{Neutrino equations of motion and system setup}
\label{sec:eom}
In this section, we introduce the equations of motion of (anti)neutrinos and describe the (anti)neutrino distributions that we use as initial conditions for our numerical calculations. 

\subsection{Equations of motion}
Under the assumption of axial symmetry, the flavor content of neutrino and antineutrino ensembles is described by density matrices of the form $\varrho_{v}(t,E, r)$ and $\bar{\varrho}_{v} (t,E,r)$, where $r$ ($v$)  is the position (velocity) along the symmetry axis, $E$ is the energy, and $t$ is time. The diagonal elements of these matrices, $\varrho_{v,\alpha \alpha}$ ($\bar{\varrho}_{v,\alpha\alpha}$), carry information on the number of (anti)neutrinos $\nu_\alpha$ ($\bar{\nu}_\alpha$) with a certain momentum; the off-diagonal elements encode the coherence between different flavors.  

In the mean-field approximation, the density matrices  evolve according to the following equations of motion: 
\begin{subequations}
\label{eq:qke-vac} 
\begin{alignat}{2}
 i(\partial_t+v\partial_r)\varrho_v(t,E, r) &=[\pm H_\mathrm{vac} + H_{\nu\nu},\varrho_v(t,E, r)]\, ,\label{eq:qke-vac1}\\
 i(\partial_t+v\partial_r)\bar{\varrho}_v(t,E, r) &=[ \mp H_\mathrm{vac} + H_{\nu\nu},\bar{\varrho}_v(t,E, r)] \, \label{eq:qke-vac2}.
 \end{alignat}
\end{subequations}
We rely on the two-flavor approximation, distinguishing between electron- and heavy-lepton-flavor (anti)neutrinos, $\nu_e$ and $\nu_x$ ($\bar{\nu}_e$ and $\bar{\nu}_x$).  The upper and lower signs in Eqs.~\ref{eq:qke-vac} correspond to the normal mass ordering (NO) and IO, respectively. The vacuum Hamiltonian $H_\mathrm{vac}$  is: 
\begin{equation}
\label{eq:H-vac}
    H_\mathrm{vac} = \frac{\omega}{2}  \left(\begin{array}{cc}-\cos 2\vartheta & \sin 2\vartheta \\ \sin 2\vartheta & \cos 2\vartheta \end{array} \right)\, ,
\end{equation}
where $\vartheta$ is the mixing angle and $\omega = \Delta m^2/(2E)$ is the oscillation frequency, which depends on the mass-squared difference $\Delta m^2 = 10^{-3}$~eV$^2$. For simplicity, we represent the matter effect via the suppressed mixing angle $\vartheta = 10^{-3}$~\cite{Hannestad:2006nj}. The neutrino-neutrino Hamiltonian is:
\begin{equation}
    H_{\nu\nu} =\mu \int \mathrm{d}E'\,\mathrm{d}v'\, (1-vv')[\varrho_{v'}(t,E', r)-\bar{\varrho}_{v^\prime}(t,E', r)]\, ,
\end{equation}
where $\mu$ is the space-averaged self-interaction strength.  

We solve Eqs.~\eqref{eq:qke-vac} in a one-dimensional box with size $L$ and periodic boundaries, assuming that all (anti)neutrinos are in flavor eigenstates at $t=0$. {We consider two fixed values of the self-interaction strength, $\mu = 63$ km$^{-1}$ and $\mu = 630$ km$^{-1}$, typical  of regions far from the core-collapse supernova core where slow flavor instabilities may develop~\cite{Duan:2006an, Fogli:2007bk}. In the low-$\mu$ case, we set $L=48\,\mathrm{km}$; in the high-$\mu$ case, $L=16\,\mathrm{km}$. In both cases, we stop the calculation at $t=L$ to avoid unphysical scenarios where (anti)neutrinos repeatedly encounter the same potential. We shorten the simulation time for $\mu =630\,\mathrm{km}^{-1}$ because fast instabilities saturate very early, so most of the evolution is limited to the growth of small-scale structures that have a negligible impact on the space-averaged density matrices $\rho_{v}(t,E) = L^{-1}\int \mathrm{d}r\, \varrho_v(t,E,r)$ and $\bar{\rho}_{v}(t,E)=L^{-1}\int \mathrm{d}r\,\bar{\varrho}_v(t,E,r)$. Since the space-averaged flavor configurations do not depend on $L$, we  choose  large values of the latter, to allow the development of slow instabilities at the edge of our parameter space  (Sec.~\ref{sec:parameter-space}).} 

{The grids for the spatial and angular coordinates, $r$ and $v$, are uniform. The former comprises $800$ bins for $r\in [0,L]$; the latter, $150$ bins for $v\in [-1,1]$. We find that changing the radial resolution by $33\%$ does not significantly affect the space-averaged distributions during the simulation time (see Appendix~\ref{sec:appendix_error} for additional details).}  The energy grid has three logarithmically spaced bins in the interval $[0.1, 1]\,\mathrm{MeV}$, $23$ linearly spaced bins for $E\in [1, 50]$~MeV, and $4$ linearly spaced bins for $E \in [50, 100]$~MeV. 

\subsection{Initial conditions}
\label{sec:parametrization}
The (anti)neutrino distributions at $t=0$, $\rho_{v, 0}(E)\equiv \rho_v(t=0,E)$ and $\bar{\rho}_{v, 0}(E)\equiv \bar{\rho}_v(t=0,E)$, are homogeneous up to random perturbations of order $10^{-3}\rho_{v,ee}$ and parameterized as follows:
\begin{subequations}
\label{eq:anisotropic_ensembles}
\begin{alignat}{2}
\rho_{v,0, \alpha\alpha}(E) &= \frac{n_{\nu_\alpha}}{K_{\nu_\alpha}}f(E,T_{\nu_\alpha})\exp\left[ -\frac{(v-1)^2}{2\sigma_{\nu_\alpha}^2}\right]\, ,\\
\bar{\rho}_{v,0, \alpha\alpha}(E) &=\frac{n_{\bar{\nu}_\alpha}}{K_{\bar{\nu}_\alpha}}f(E,T_{\bar{\nu}_\alpha}) \exp\left[ -\frac{(v-1)^2}{2\sigma_{\bar{\nu}_\alpha}^2}\right] \, ,
\end{alignat}
\end{subequations}
where $f(E,T)$ is the dimensionless Fermi-Dirac function: 
\begin{equation}
\label{eq:fermi-dirac}
    f(E,T_{\nu_\alpha}) = \frac{(E/T_{\nu_\alpha})^2}{\exp (E/T_{\nu_\alpha})+1}\, .
\end{equation}
 The normalization factors $K_{\nu_\alpha}$ and $K_{\bar{\nu}_\alpha}$ are such that $\int \mathrm{d}E\,\mathrm{d}v\,\rho_{v,0,\alpha\alpha}=n_{\nu_\alpha}$ and $\int \mathrm{d}E\,\mathrm{d}v\,{\bar{\rho}}_{v, 0,\alpha\alpha} =n_{\bar{\nu}_\alpha}$. The parameter $n_{\nu_\alpha}$ ($n_{\bar{\nu}_\alpha}$) is the density of $\nu_\alpha$ ($\bar{\nu}_\alpha$)  divided by the all-flavor neutrino density, which is given by $\mu  / (\sqrt{2}G_\mathrm{F})$ and varies as a free parameter. Thus, the densities of $\nu_e$ and $\nu_x$ are related by $n_{\nu_e}+n_{\nu_x}=1$. 

The stability of the ensemble depends on  
$12$ variables: $\lbrace \mu,\,T_{\nu_\alpha},\,T_{\bar{\nu}_\alpha},\,n_{\nu_e},\,n_{\bar{\nu}_\alpha},\,\sigma_{\nu_\alpha}^2,\, \sigma^2_{\bar{\nu}_\alpha}\rbrace$. To reduce the number of free parameters, we limit our investigation to cases with identical $\nu_x$ and $\bar{\nu}_x$ distributions and impose  $\sigma^2_{\bar{\nu}_e}=\sigma^2_{\nu_x}$. Moreover, we consider  $\sigma_{\nu_e}^2>\sigma_{\bar{\nu}_e}^2$ and $T_{\nu_x}>T_{\bar{\nu}_e}>T_{\nu_e}$, as expected in astrophysical settings~\cite{Tamborra:2017ubu,Brandt:2010xa,Keil:2002in,Mirizzi:2015eza}. 

The inequalities $T_{\nu_x}>T_{\nu_e,\bar\nu_e}$ and $n_{\nu_x}<n_{\nu_e}$ guarantee the existence of $\nu_e$--$\nu_x$ and $\bar\nu_e$--$\bar\nu_x$ spectral crossings. {We stress that these conditions are not sufficient for slow conversion: in general, assessing the stability of the ensemble would require looking for complex solutions of the neutrino dispersion relation ~\cite{Dasgupta:2009mg,Banerjee:2011fj, Dasgupta:2021gfs,DedinNeto:2023ykt,Fiorillo:2024pns}. However, in this study, we focus on  regions of the parameter space where slow instabilities do occur (see Sec.~\ref{sec:parameter-space})}. 

\begin{figure}[t]
    \centering
    \includegraphics[width=\linewidth]{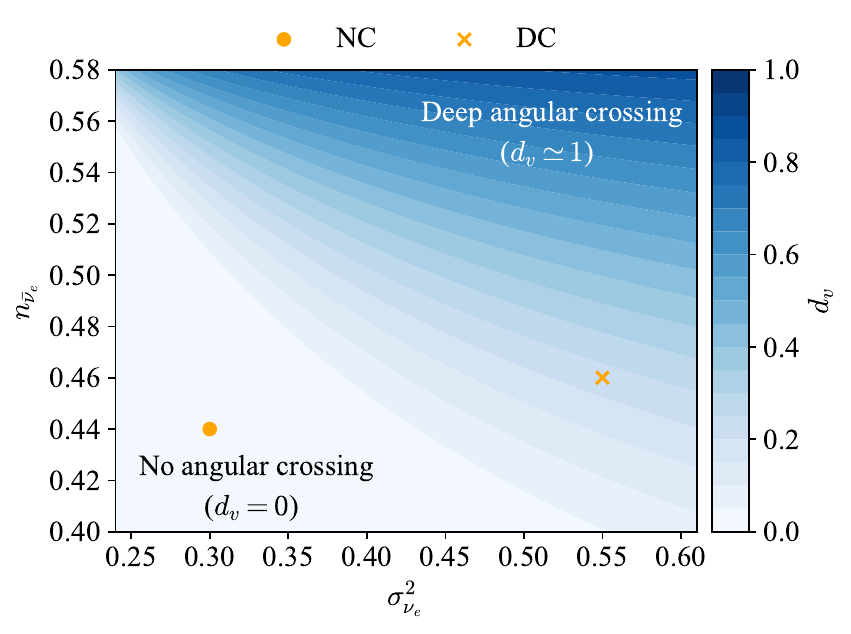}
    \caption{Characterization of the (anti)neutrino  parameter space according to the depth  of the angular crossing in the energy-integrated ELN-XLN distribution ($d_v$, cf.~Eq.~\ref{eq:crossing_depth}), for $n_{\nu_e}=0.60$. All configurations in this plane have $\nu_e$--$\nu_x$ and $\bar{\nu}_e$--$\bar{\nu}_x$ spectral crossings, necessary for slow collective instabilities.  Ensembles with high $\sigma_{\nu_e}^2$ and high  $n_{\bar{\nu}_e}$ have  deep $\nu_e$--$\bar{\nu}_e$ angular crossings as well, so  fast conversion is expected. The orange markers show the distributions adopted in NC (No Crossing, with $d_v=0$)  and  DC (Deep Crossing, with $d_v\sim 0.2$) shown in Figs.~\ref{fig:approximation_slow} and \ref{fig:approximation_fast}, respectively.
    }
     \label{fig:parameter_space}
\end{figure}

{A subset of the ensembles in our parameter space also have crossings in the energy-integrated ELN-XLN distribution, so they are prone to fast conversion} regardless of their energy spectra~\cite{Morinaga:2021vmc}. To gain insight into the type of flavor instability that could dominate each region of our parameter space for a given $\nu_e$ density ($n_{\nu_e}$), Fig.~\ref{fig:parameter_space} classifies the angular distributions in Eqs.~\eqref{eq:anisotropic_ensembles} according to the depth of their angular crossing, which we define as:
\begin{equation}
\label{eq:crossing_depth}
    d_v = \frac{\int \mathrm{d}v\,\Theta (\int \mathrm{d}E\,[\bar{\rho}_{v,0,ee}-{\rho}_{v,0,ee}])\int\,\mathrm{d}E\, [\bar{\rho}_{v,0,ee}-{\rho}_{v,0,ee}]}{\int \mathrm{d}v\,\Theta (\int \mathrm{d}E\,[{\rho}_{v,0,ee}-\bar{\rho}_{v,0,ee}])\int\,\mathrm{d}E\, [\rho_{v,0,ee}-\bar{\rho}_{v,0,ee}]}\, .
\end{equation}
Fast conversion drives the flavor evolution of (anti)neutrino distributions with high $\sigma_{\nu_e}^2$ and/or $n_{\bar{\nu}_e}$;  distributions with small $\sigma_{\nu_e}^2$ and/or $n_{\bar{\nu}_e}$ ($d_v=0$) may only undergo slow flavor conversion. {Between these two extremes, the interplay between the vacuum mixing and self-interactions can affect the quasi-steady state of ensembles with shallow angular crossings~\cite{DedinNeto:2023ykt}.}

\section{Forecast of the steady-state flavor configuration}
\label{sec:ansatz}
In this section, we provide semi-analytic approximations of the quasi-steady state configuration achieved by (anti)neutrino ensembles undergoing slow and fast flavor instabilities. 
First, we compare our semi-analytic ansatz  with the numerical solutions of Eqs.~\eqref{eq:qke-vac} for two selected initial configurations, with model parameters fixed as in Table~\ref{tab:parameters} and highlighted in Fig.~\ref{fig:parameter_space}: NC (with $d_v=0$) and DC (with $d_v\sim 0.2$). Then,  we undertake  a more general exploration of the (anti)neutrino parameter space.

\begin{figure*}
    \centering
    \includegraphics[width=\linewidth]{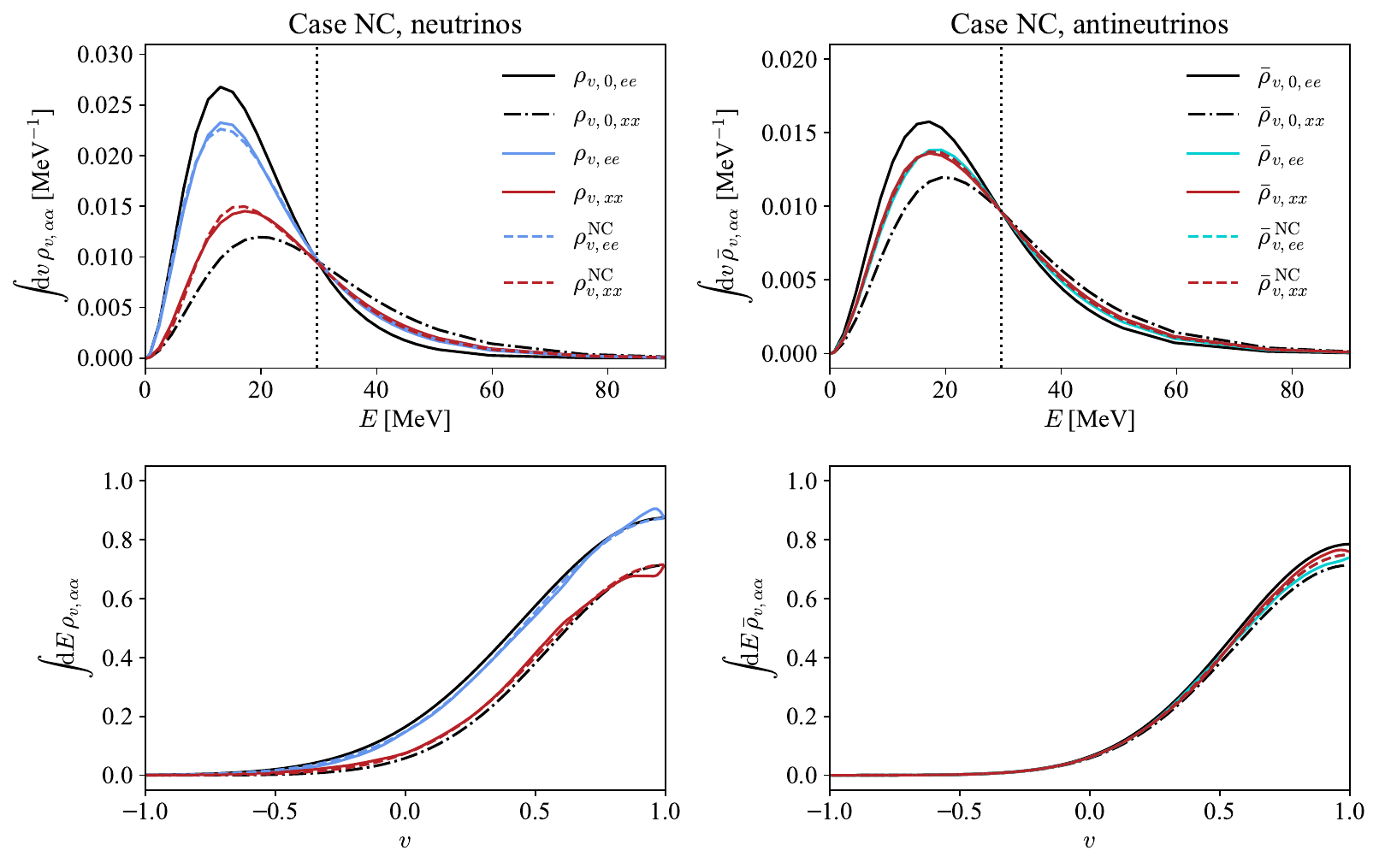}
    \caption{Energy (top panels, after angle integration) and angular (bottom panels, after energy integration) flavor distributions for the ensemble NC, defined by the parameters in Table~\ref{tab:parameters} (cf.~also  Fig.~\ref{fig:parameter_space}), before (black lines) and after (lines in color) flavor conversion for neutrinos (left) and antineutrinos (right). The dotted vertical lines mark the energy crossings, whereas the NC distributions do not have any angular crossings. The  solid  lines show the numerical solution of Eqs.~\eqref{eq:qke-vac} in the quasi-steady state, at $t = L$; the dashed lines represent the semi-analytic ansatz in Eqs.~\eqref{eq:approximation_slow_ee}--\eqref{eq:approximation_slow_xx} and agree with the numerical solution to within $5\%$ relative error (see Sec.~\ref{sec:parameter-space}). 
     Antineutrinos reach  equipartition across all energies; neutrinos achieve flavor equipartition only in the high-energy tail of the spectrum. We show the quasi-steady flavor configuration in NO for illustrative purposes. }
    \label{fig:approximation_slow}
\end{figure*}

\subsection{Slow flavor conversion ($d_v=0$)}
\label{sec:ansatz_slow}

Figure~\ref{fig:approximation_slow} shows the initial and quasi-steady state angle-integrated energy spectra (top panels) and energy-integrated angular distributions (bottom panels) of the ensemble NC with parameters fixed as  in Table~\ref{tab:parameters} (see also Fig.~\ref{fig:parameter_space}). The initial configuration (black lines) has $e$-$x$ spectral crossings in both the neutrino (left) and antineutrino (right) spectra, but no ELN-XLN angular crossing; hence, we expect slow instabilities to drive flavor conversion. We assume NO  for illustrative purposes (flavor conversion in IO  is  suppressed, as discussed in Sec.~\ref{sec:variations}).

\begin{table}[t]
 \caption{Parameters used in the numerical examples of Secs.~\ref{sec:ansatz_slow} and \ref{sec:ansatz_fast}; see Fig.~\ref{fig:parameter_space}.} 
    \label{tab:parameters}
    \centering
    \begin{tabular}{c|ccccccc}
    \toprule
      Case & $\mu$ [km$^{-1}$] &$\sigma_{\nu_e}^2$ & $\sigma_{\bar{\nu}_e}^2$ & $\lbrace T_{\nu_e},T_{\bar{\nu}_e},T_{\nu_x}\rbrace$ [MeV] &  $n_{\nu_e}$ & $n_{\bar{\nu}_e}$\\\midrule
      NC & 63 & 0.30 & 0.20 & $\lbrace 6.0, 7.5, 9.0\rbrace$ & 0.60 & 0.44 \\ 
      DC & 630 & 0.55 & 0.20 & $\lbrace 6.0, 7.5, 9.0\rbrace$ & 0.60 & 0.46 \\\bottomrule
    \end{tabular}
\end{table}

The solid lines in color represent the  solution of Eqs.~\eqref{eq:qke-vac} at $t=L$.  In the neutrino sector (bottom left), interactions of the form $
\nu_e(E<E_c,v)+\nu_x(E'>E_c,v')\to \nu_e(E',v')+\nu_x(E,v)$ tend to equalize  the neutrino spectra for $E>E_c$.  
Interactions of $\nu_e$--$\bar{\nu}_e$ pairs further shape the energy spectra for $E<E_c$. As illustrated in the bottom-right panel of Fig.~\ref{fig:approximation_slow}, $\bar{\nu}_e$--$\bar{\nu}_x$ interactions drive the angle-integrated antineutrino spectra towards equipartition.

To approximate the quasi-steady configuration of the NC ensemble, we assume equipartition for all antineutrinos and for neutrinos with $E>E_c$. Then, we  rescale the part of the neutrino spectra with $E<E_c$ by an angle-independent factor: 
\begin{align}
\rho_{v,ee}^\mathrm{NC}(E)&= \left\lbrace \begin{array}{rl} 
       [R^\mathrm{NC}_{ee}\rho_{v,0,ee} +\tilde{\rho}_{v,0,xx}](E)& (E<E_c)\,\\
       \displaystyle {\frac{1}{2}\mathrm{Tr}[\rho_{v,0}(E)]} &(E\geq E_c)\,\end{array}\right.\, , \label{eq:approximation_slow_ee}\\ 
\bar{\rho}_{v,ee}^\mathrm{\,NC}(E) 
 &= \displaystyle\frac{1}{2}\mathrm{Tr}[\bar{\rho}_{v,0}(E)]  \, ,\label{eq:approximation_slow_eb}\\
\rho_{v,xx}^\mathrm{NC}(E) &=
\rho_{v,0,xx}(E) + [\rho_{v,0,ee}(E)-\rho^\mathrm{NC}_{v,ee}(E)]\, ,\\
\bar{\rho}_{v,xx}^\mathrm{\,NC}(E) &=\bar{\rho}_{v,0,xx}(E) + [\bar{\rho}_{v,0,ee}(E)-\bar{\rho}^\mathrm{\,NC}_{v,ee}(E)]\, .\label{eq:approximation_slow_xx}
 \end{align}
Here, $\tilde{\rho}_{v,0,xx}(E) \equiv [1-R_{xx}^\mathrm{NC}(E)]S_x(E)G_e(v)$, where $S_{\alpha}(E) \equiv \int \mathrm{d}v'\, \rho_{v',0,\alpha\alpha}(E)$ is the initial $\nu_\alpha$ energy spectrum and $G_\alpha(v)\equiv( n_{\nu_\alpha})^{-1}\int \mathrm{d}E'\,\rho_{v,0,\alpha\alpha}(E')$ is the $\nu_\alpha$ angular distribution, normalized so that $\int \mathrm{d}v\, G_{\alpha}(v) =1$.
The function $R^{\mathrm{NC}}_{\alpha\alpha}$ represents the fraction of $\nu_\alpha$ that does not change flavor:
\begin{equation}
\label{eq:survival_slow}
    R_{\alpha\alpha}^\mathrm{NC}(E) = a_\alpha \frac{f(E,\langle T \rangle)}{\int \mathrm{d}E'\, f(E',\langle T \rangle)}\, ,
\end{equation}
where $f(E,\langle T \rangle)$ is defined as in Eq.~\eqref{eq:fermi-dirac} and 
$\langle T\rangle  = (T_{\nu_e}+T_{\nu_x})/2$. The parameters $a_e$ and $a_x$ in Eq.~\eqref{eq:survival_slow} are determined by the requirement that the angle-integrated energy spectrum be continuous at $E=E_c$: 
\begin{equation}
    \lim_{E\to E_c^{-}}\int \mathrm{d}v\,\rho_{v,ee}^\mathrm{NC}(E) = \frac{1}{2}\int \mathrm{d}v\,\mathrm{Tr}[\rho_{v,0}(E_c)]\, ,
\end{equation}
which imposes $a_e=a_x\equiv a$. Lepton-number conservation then determines $a$ for each set of initial conditions:
\begin{equation}
\label{eq:eln-conservation}
    \int \mathrm{d}E\,\mathrm{d}v\,[(\rho_{v,ee}^\mathrm{NC}-\bar{\rho}_{v,ee}^\mathrm{\,NC})-(\rho_{v,0,ee}-\bar{\rho}_{v,0,ee})](E) =0\,.
\end{equation}

Equations~\eqref{eq:approximation_slow_ee}-\eqref{eq:approximation_slow_xx} are plotted with  dashed lines in Fig.~\ref{fig:approximation_slow}. Our ansatz successfully reproduces the angle-integrated energy spectra obtained by solving Eqs.~\eqref{eq:qke-vac}, but the assumption that the survival ratios are angle-independent does not capture the increase in the $\nu_e$ density at $v\gtrsim 0.9$ in the top panel of Fig.~\ref{fig:approximation_slow}. This implies that equipartition is only an angle-averaged result; in this case, $\nu_x$ traveling on-axis ($v\simeq 1$) convert into $\nu_e$ more  than those with  $v \lesssim 1$.

\begin{figure*}
    \centering
    \includegraphics[width=\linewidth]{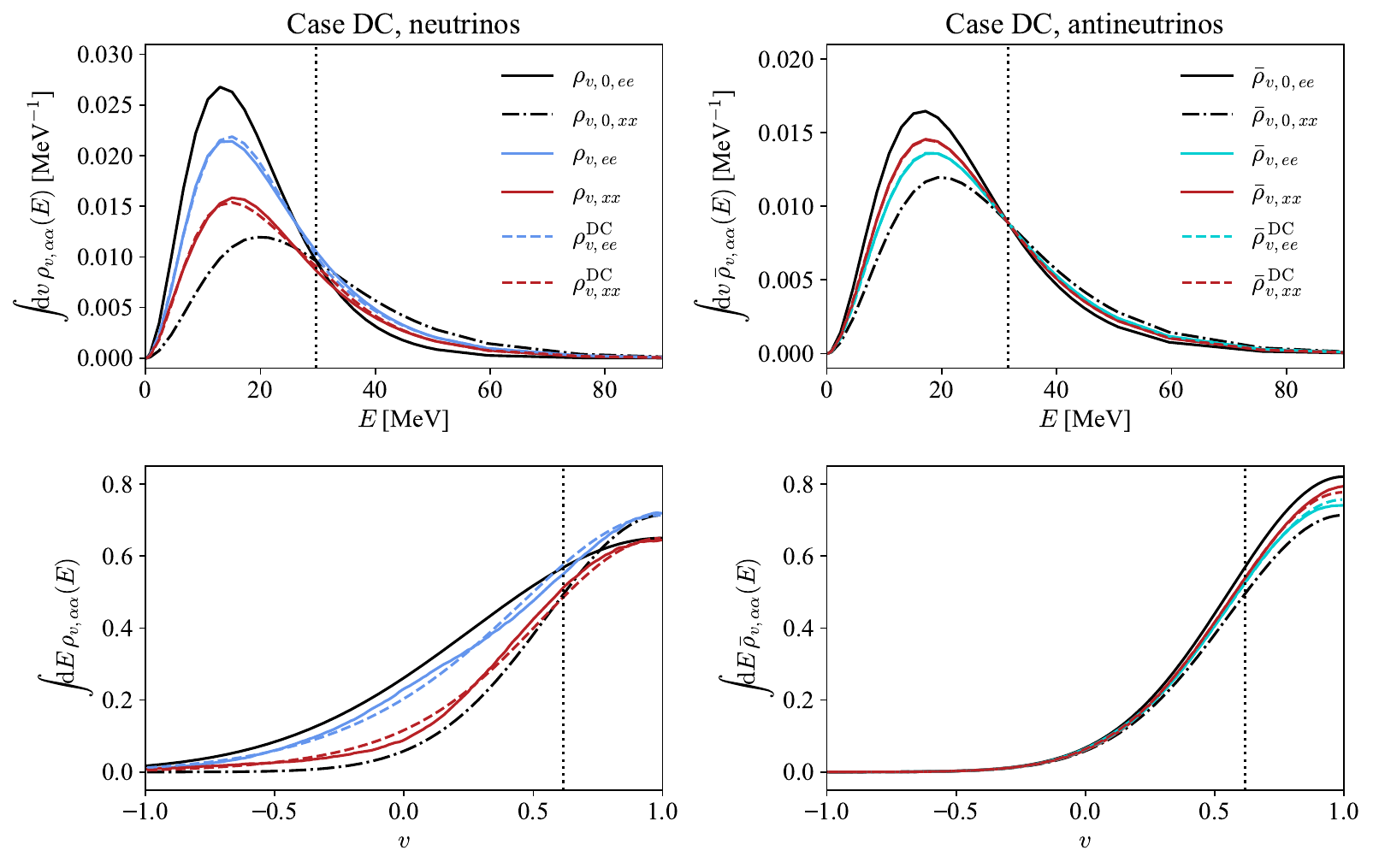}
     \caption{Same as Fig.~\ref{fig:approximation_slow}, but for the ensemble DC (see Fig.~\ref{fig:parameter_space} and Table~\ref{tab:parameters}), which has an ELN-XLN crossing with $d_v\sim 0.2$. The dotted vertical lines mark the initial energy (top panels) and $\nu_e\bar{\nu}_e$ angular crossings (bottom panels). The semi-analytic approximations (dashed lines in color) follow  Eqs.~\eqref{eq:approximation_fast_ee}--\eqref{eq:approximation_fast_xx}. 
     The survival fraction in Eq.~\eqref{eq:fast_survival} matches the antineutrino distribution almost exactly; as for  neutrinos, the relative error between the numerical solution and  the semi-analytic approximation is $\sim10\%$ (see Sec.~\ref{sec:parameter-space}).}
    \label{fig:approximation_fast}
\end{figure*}

\subsection{Fast flavor conversion ($d_v>0$)}
\label{sec:ansatz_fast}

When the energy-integrated ELN-XLN distribution has an angular crossing, the ensemble is unstable regardless of $e$--$x$ spectral crossings. If fast  conversion dominates the evolution, the negative part of the energy-integrated ELN-XLN distribution tends to change sign, while conserving the lepton number and the overall number of particles. We modeled such a quasi-steady state in Ref.~\cite{Goimil-Garcia:2025ozm}, focusing  on  single-energy ensembles and assuming an initial flavor configuration composed of electron (anti)neutrinos only. 
Here, we generalize our earlier findings to account for spectral (anti)neutrino distributions as well as non-oscillated configurations made of $e$ and $x$ flavors. 

Figure~\ref{fig:approximation_fast} shows the quasi-steady (anti)neutrino configuration of the ensemble DC (see Fig.~\ref{fig:parameter_space} and Table~\ref{tab:parameters}) in NO. The self-interaction strength $\mu =630~$km$^{-1}$ is lower than expected in  core-collapse supernovae in the regions where angular crossings first develop, see e.g.~Refs.~\cite{Capozzi:2020syn,Nagakura:2019sig, Nagakura:2021hyb, Cornelius:2025tyt,  Cornelius:2025nvd}. However, it is still $10^3$ times higher than the average  frequency of vacuum oscillations, 
so the impact of the vacuum term on the quasi-steady state is expected to be small (see Sec.~\ref{sec:parameter-space}).

Flavor conversion is driven by the angular crossing in the energy-integrated ELN-XLN distribution (bottom panels, black lines) and results in a net increase of $x$-flavor (anti)particles via $\nu_e\bar{\nu}_e\to \nu_x\bar{\nu}_x$; however, the direction of flavor conversion is both angle- and energy-dependent. The densities of $\nu_e$ and $\bar{\nu}_e$ increase at high energies (top panels, solid  lines), because  $T_{\nu_x}>T_{\bar{\nu}_e}>T_{\nu_e}$ and $n_{\bar{\nu}_e}>n_{\bar{\nu}_x}$ result in $e$--$x$ spectral crossings. For $\nu_e$, this increase is localized in the angular range with $v\gtrsim 0.7$ (bottom panel, solid  lines), where $\int \mathrm{d}E\,\rho_{v,0, xx}(E)>\int \mathrm{d}E\,\rho_{v,0,ee}(E)$; for $v\lesssim 0.7$, the $\nu_e$ density decreases to preserve the lepton number.

In order to model the quasi-steady flavor configuration, we assume that the    fraction of $\bar{\nu}_e$'s that survives is described by the following empirical expression:
\begin{equation}
\label{eq:fast_survival}
        R= \frac{1}{3}\left[1 +\frac{1-n_{\nu_e}}{2-(n_{\nu_e}-n_{\bar{\nu}_e})} \right]\, ,
\end{equation}
which accounts for the fact that fast instabilities can overshoot flavor equipartition ($R<0.5$) when $n_{\nu_e}$ is large  and the ELN is low (cf.~Sec.~\ref{sec:parameter-space}).
This is consistent with the single-energy predictions of  Ref.~\cite{Goimil-Garcia:2025ozm} for $n_{\nu_e}=1$.

For neutrinos, we apply Eq.~\eqref{eq:fast_survival} above the energy threshold $E_\mathrm{th} = 5.4 T_{\nu_e}$; for  $E<E_{\rm{th}}$, we write the $\nu_e$ and $\nu_x$ distributions as  linear combinations of $G_e(v)$ and $G_x(v)$:
\begin{align}
       \rho_{v,ee}^\mathrm{DC}(E)  &= \left\lbrace 
        \begin{array}{ll} 
          \displaystyle F(E,\mathcal{T})\sum_\alpha b_\alpha G_\alpha(v) & (E\leq E_{\rm{th}})\,,\, \\
         \left[R\rho_{v,0,ee}+R'\rho_{v,0,xx}\right](E) 
         & (E> E_{\rm{th}})\,,\
        \end{array} \right.\label{eq:approximation_fast_ee}\\ 
        \bar{\rho}_{v,ee}^\mathrm{\,DC}(E) &= R \bar{\rho}_{v,0,ee}(E)+R'\bar{\rho}_{v,0,xx}(E)\,,\label{eq:approximation_fast_eb}\\
    \rho_{v,xx}^\mathrm{DC}(E)  &=
\rho_{v,0,xx}(E) + [\rho_{v,0,ee}(E)-\rho_{v,ee}^\mathrm{DC}(E)]\, ,\\
\bar{\rho}_{v,xx}^\mathrm{\,DC}(E)  &=
\bar{\rho}_{v,0,xx}(E) + [\bar{\rho}_{v,0,ee}(E)-\bar{\rho}^\mathrm{\, DC}_{v,ee}(E)]\,  .\label{eq:approximation_fast_xx}
    \end{align}
Here, $\alpha=e$ or $x$, 
$R'\equiv 1-R$ and $F(E,T)\equiv f(E,T)/\int \mathrm{d}E'\, f(E',T)\Theta(E_{\rm{th}}-E')$, where $f(E,T)$ is defined as in Eq.~\eqref{eq:fermi-dirac}. 
To determine the constants $b_\alpha$, we assume that fast conversion for $v=1$ can either lead to equipartition or swap the neutrino distributions (cf. bottom left panel of Fig.~\ref{fig:approximation_fast}, solid  lines in color): 
\begin{equation}
\label{eq:fast_eqv1}
    \begin{array}{rl}
   & \displaystyle \sum_{\alpha =e,x}b_\alpha G_\alpha (v=1)+\displaystyle\int_{E_{\rm{th}}}^\infty \mathrm{d}E\,\rho_{v=1,ee}^\mathrm{DC}(E) \\
   &\quad = \displaystyle\mathrm{max}\left\lbrace\int \mathrm{d}E\, \frac{\mathrm{Tr}[\rho_{v=1,0}]}{2}\,,\, \int \mathrm{d}E\, \rho_{v=1,0,xx} \right\rbrace\, ,
    \end{array}
\end{equation}
and  impose lepton-number conservation as in Eq.~\eqref{eq:eln-conservation}. 
The requirement that the angle-integrated spectra should be continuous at $E=E_{\rm{th}}$ implicitly defines $\mathcal{T}$, determining the solution for  $E\leq E_{\rm{th}}$.

The semi-analytic approximations in Eqs.~\eqref{eq:approximation_fast_ee}--\eqref{eq:approximation_fast_xx} are plotted with   dashed lines in Fig.~\ref{fig:approximation_fast}. The effective temperature $\mathcal{T}$ slightly overestimates (underestimates) the $\nu_e$ ($\nu_x$) density at $E\simeq 30$~MeV (top left panel), and Eq.~\eqref{eq:approximation_fast_eb} predicts less flavor conversion in the forward direction than actually observed (bottom right). However,  the agreement between the analytic and the  numerical solutions is overall good. Notably, the survival probability $R$ defined in Eq.~\eqref{eq:fast_survival} describes the high-energy tail of the neutrino spectra more accurately than equipartition. 

\subsection{Flavor conversion in the parameter space of angular distributions}
\label{sec:parameter-space}

\begin{figure}[t]
    \centering
    \includegraphics[width=\linewidth]{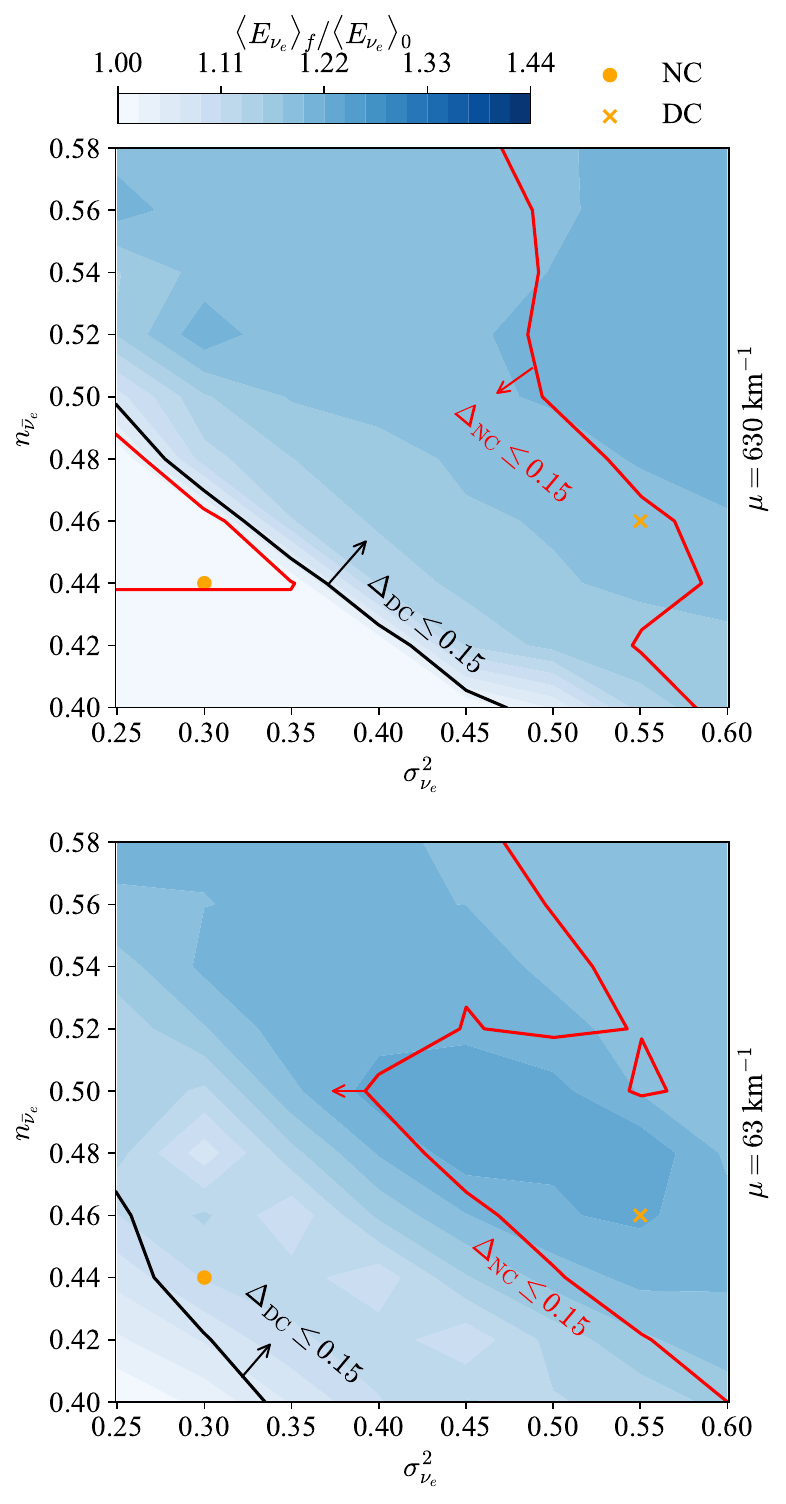}
    \caption{Comparison between the numerical solution of Eqs.~\eqref{eq:qke-vac} and the semi-analytic approximations of Secs.~\ref{sec:ansatz_slow} and \ref{sec:ansatz_fast} in the plane of  Fig.~\ref{fig:parameter_space}, assuming NO  and using the parameters $\lbrace T_{\nu_l},\,T_{\bar{\nu}_e},\,\sigma^2_{\bar{\nu}_e},n_{\nu_e}\rbrace$ in Table~\ref{tab:parameters}. The heat map shows the evolution of the average $\nu_e$ energy, to quantify the amount of flavor conversion. The orange markers point at the NC (dot) and DC (cross) ensembles in Table~\ref{tab:parameters}. The red (black) contours mark the regions where the relative error between the numerical and the semi-analytic approximation for slow (fast) flavor conversion is $\leq 0.15$, according to Eq.~\eqref{eq:error}.}
    \label{fig:app_error}
\end{figure}

We now explore flavor conversion for a broader suite of initial conditions and assess the range of validity of Eqs.~\eqref{eq:approximation_slow_ee}--\eqref{eq:approximation_slow_xx} and Eqs.~\eqref{eq:approximation_fast_ee}--\eqref{eq:approximation_fast_xx}. Using the values of $n_{\nu_e}$,  $\sigma_{\bar{\nu}_e}^2$ and $\mu$ from Table~\ref{tab:parameters} and assuming NO, we solve Eqs.~\eqref{eq:qke-vac} in the  ($\sigma_{\nu_e},n_{\bar{\nu}_e}$) plane of Fig.~\ref{fig:parameter_space} for $\lbrace T_{\nu_e},\,T_{\bar{\nu}_e},\,T_{\nu_x}\rbrace =\lbrace 6,\,7.5,\,9\rbrace$~MeV, which have been  used for the NC and DC configurations.

To quantify the magnitude of flavor conversion  across the parameter space  in Fig.~\ref{fig:parameter_space}, we introduce  the ratio between the average energy of $\nu_e$ at the end of the simulation ($\langle E_{\nu_e}\rangle(t=L)\equiv \langle E_{\nu_e}\rangle_f$) and the one at the beginning ($\langle E_{\nu_e}\rangle(t=0)\equiv \langle E_{\nu_e}\rangle_0$), with:
\begin{equation}
\label{eq:avg-energy}
    \langle E_{\nu_e} \rangle (t)= \frac{\int \mathrm{d}v\,\mathrm{d}E\,E\rho_{v,ee}(t,E)}{\int \mathrm{d}v\,d\mathrm{E}\,\rho_{v,ee}(t,E)}\, .
\end{equation}
For ensembles without angular crossings ($d_v =0$,  bottom-left corner of Fig.~\ref{fig:parameter_space}), the average energy  provides a better indication of the amount of flavor conversion  than the energy- and angle-integrated $\nu_e$ survival probability. The latter is conserved by  $\nu_e$-$\nu_x$ and $\bar{\nu}_e$-$\bar{\nu}_x$ interactions that dominate the evolution, but $\langle E_{\nu_e}\rangle_f/\langle E_{\nu_e}\rangle_0$ increases because $T_{\nu_x}>T_{\nu_e}$.

Figure~\ref{fig:app_error} shows isocontours of $\langle E_{\nu_e}\rangle_f/\langle E_{\nu_e}\rangle_0$ in the plane spanned by $\sigma^2_{\nu_e}$ and $n_{\bar\nu_e}$ (cf.~Fig.~\ref{fig:parameter_space} for the related isocontours of $d_v$) for $\mu = 630$~km$^{-1}$ (top panel) and $\mu = 63$~km$^{-1}$ (bottom panel). Regardless of $\mu$, the amount of flavor conversion is greater in the top right corner of the parameter space, where fast instabilities drive flavor conversion. The higher self-interaction strength enhances flavor conversion around the peak of the energy distribution, but suppresses the cascade of flavor conversion towards small angular scales caused by the vacuum term and advection. However, for these ensembles, comparing the top and bottom panels of Fig.~\ref{fig:app_error} shows that $\langle E_{\nu_e}\rangle_f$ changes mildly with $\mu$.

In configurations with more shallow angular crossings (center right of the parameter space), the interplay between the vacuum term and neutrino-neutrino interactions enhances flavor conversion in the $\mu = 63\,$km$^{-1}$ case. This effect becomes negligible as the separation of scales between the self-interaction strength and the typical vacuum frequency increases. Similarly, ensembles without angular crossings (lower-left corner of the parameter space) approach stability when $\mu =630$km$^{-1}$, i.e., $\langle E_{\nu_e}\rangle_f\to \langle E_{\nu_e}\rangle_0 $  with increasing $\mu$. 

Our semi-analytic neutrino distributions depend on the antineutrino ones via lepton-number conservation (cf. Eq.~\ref{eq:eln-conservation}), so we focus on the agreement between $\rho_{v,ee}^\mathrm{NC,DC}(E)$ and the numerical solution of Eqs.~\eqref{eq:qke-vac} as an overall assessment of the ansatz. To quantify the error of Eqs.~\eqref{eq:approximation_slow_ee}--\eqref{eq:approximation_slow_xx} and Eqs.~\eqref{eq:approximation_fast_ee}--\eqref{eq:approximation_fast_xx}, we compute:
\begin{equation}
    \label{eq:error}
    \Delta_{\mathrm{NC,\,DC}} =\frac{1}{n_{\nu_e}}\int \mathrm{d}v\,\mathrm{d}E\, \left|\rho_{v,ee}(t=L,E)-\rho_{v,ee}^{\mathrm{NC,\,DC}}(E) \right|\,.
\end{equation}

The red (black) contours of Fig.~\ref{fig:app_error} show the regions with $\Delta_\mathrm{NC}\leq 0.15$ ($\Delta_{\mathrm{DC}}\leq 0.15$) in the plane of angular distributions. There is a region of overlap between the approximations for slow and fast conversion, where Eqs.~\eqref{eq:approximation_slow_ee}--\eqref{eq:approximation_slow_xx} and Eqs.~\eqref{eq:approximation_fast_ee}--\eqref{eq:approximation_fast_xx} both match  the numerical solution of Eqs.~\eqref{eq:qke-vac} within $15\%$ error.
The NC ansatz generally agrees with the numerical results when the $\nu_e$ and $\nu_x$ distributions have similar widths, but fails in the right part of  our parameter space ($\sigma^2_{\nu_e}\gg \sigma_{\nu_x}^2$). Flavor conversion strongly distorts the shape of the angular distributions in this region, so the ansatz $\rho_{v,ee}^\mathrm{NC}(E<E_c)\sim G_{e}(v)$ is too simple to capture the numerical result. Conversely, the DC ansatz overpredicts the amount of flavor conversion in the forward direction for the ensembles in the lower-left corner of our parameter space, where the distributions do not have an angular crossing. Moreover, both the NC and DC prescriptions assume that the fraction of (anti)neutrinos  that changes flavor does not depend on their propagation angle ($v$). Although $\Delta_{NC,\,DC}$ is generally small, flavor conversion is stronger around the peak of the angular distributions, where (anti)neutrinos are more abundant. This angular dependence is especially relevant for distributions on the verge of developing an ELN-XLN crossing or with shallow crossings, where the interplay between the self-interaction  and the vacuum terms can lead to overconversion with respect to our prediction. 

The NC and DC approximations reproduce angle-integrated  spectra with comparable accuracy. The continuity condition for Eq.~\eqref{eq:approximation_fast_ee} generally leads to $\mathcal{T}> \langle T\rangle $ so $\rho^\mathrm{DC}_{v,\alpha\alpha}(E)$ tends to have a larger average energy than $\rho^\mathrm{NC}_{v,\alpha\alpha}(E)$ when applied to the same set of  initial conditions. The numerical antineutrino distributions typically lie between the semi-analytic prescriptions provided in Eqs.~\eqref{eq:approximation_slow_eb} and \eqref{eq:approximation_fast_eb}, closer to the former in the lower-left region ($d_v=0$) and to the latter in the upper-right part ($d_v\sim 1$) of our parameter space. The error in the antineutrino sector follows similar trends to $\Delta_{\mathrm{NC, DC}}$, but is generally lower because the initial distributions of $\bar{\nu}_e$ and $\bar{\nu}_x$ are more similar to each other than those of $\nu_e$ and $\nu_x$.

\section{Energy dependence and mass ordering}
\label{sec:variations}
In this section, we investigate the dependence of the quasi-steady flavor configuration on the choice of neutrino temperatures and mass ordering.  

\subsection{Spectral shape}
\label{sec:temperature}
\begin{figure}[t]
\centering
\includegraphics[width=\linewidth]{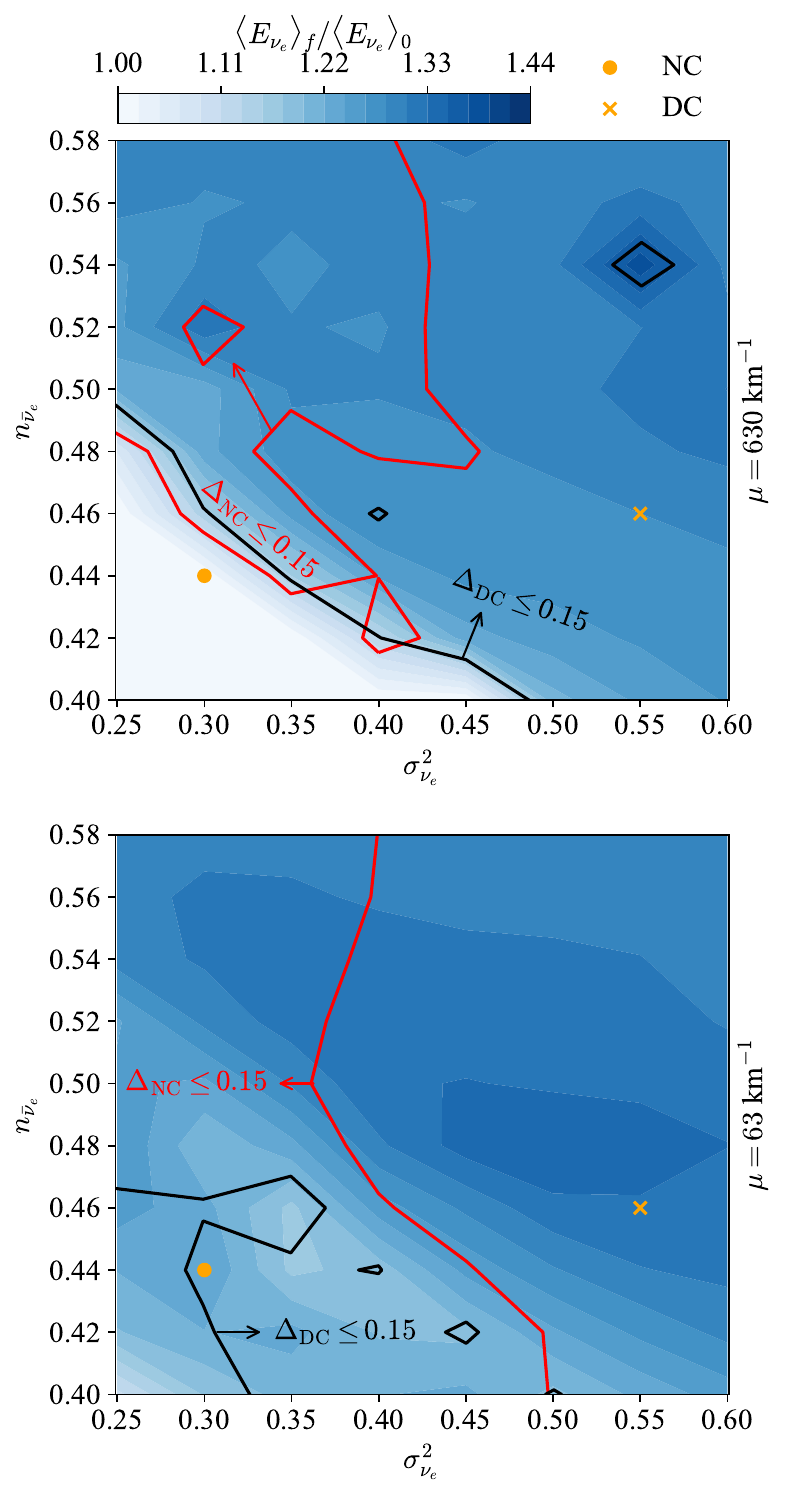}
\caption{Analogous of Fig.~\ref{fig:app_error}, but for the neutrino temperatures $\lbrace T_{\nu_e},\,T_{\bar{\nu}_e},\,T_{\nu_x}\rbrace =\lbrace 4,\,5.5,\,7\rbrace$~MeV. 
The ratio  $\langle E_{\nu_e}\rangle_f/\langle E_{\nu_e}\rangle_0$ is overall larger than in Fig.~\ref{fig:app_error} due to enhanced flavor conversion for $E \gtrsim E_c$. Our semi-analytic prescriptions reproduce well the quasi-steady obtained by solving the neutrino kinetic equations numerically.}
\label{fig:app_error_T4T55T7}
\end{figure}

Figure~\ref{fig:app_error_T4T55T7} shows the ratio between the initial average  energy of $\nu_e$ and the one  in the quasi-steady state for ensembles with the same energy-integrated angular distributions as those in Fig.~\ref{fig:app_error}, but with lower initial temperatures: $\lbrace T_{\nu_e},\,T_{\bar{\nu}_e},\,T_{\nu_x}\rbrace =\lbrace 4,\,5.5,\,7\rbrace$~MeV. Flavor conversion follows the same qualitative trends described in Secs.~\ref{sec:ansatz_slow} and \ref{sec:ansatz_fast}, but $\nu_x\to \nu_e$ ($\bar{\nu}_x\to \bar{\nu}_e$) reactions are enhanced at high energies, because the larger relative difference between the (anti)neutrino temperatures is responsible for  steeper energy crossings. The increase in flavor mixing at $E>E_c$ is responsible for the higher ratio of $\langle E_{\nu_e}\rangle_f/\langle E_{\nu_e}\rangle_0$ visible everywhere in our parameter space.

Since the ratio $T_{\nu_x}/T_{\nu_e}$ is greater, the distinction between equipartition and Eq.~\eqref{eq:fast_survival} is more pronounced than in Sec.~\ref{sec:parameter-space}, reducing the region of overlap between the prescriptions in Eqs.~\eqref{eq:approximation_slow_ee}--\eqref{eq:approximation_slow_xx} and \eqref{eq:approximation_fast_ee}--\eqref{eq:approximation_fast_xx}.
Our approximations  still match the numerical ones in most of the parameter space, but overall the error is slightly larger. As noted in Sec.~\ref{sec:ansatz_slow}, $\nu_x\to \nu_e$ conversion is anisotropic, with larger flavor mixing taking place in the proximity of $v\sim 1$, where the density of particles is the largest. Hence, the numerical distributions can be more forward-peaked than predicted by Eqs.~\eqref{eq:approximation_slow_ee} and \eqref{eq:approximation_fast_ee}. 

The vacuum frequency $\omega$ is higher on average for this set of neutrino temperatures, so the interplay between the vacuum term and neutrino-neutrino interactions tends to be  stronger than in Sec.~\ref{sec:parameter-space}. A comparison between the top and bottom panels  of Fig.~\ref{fig:app_error_T4T55T7} suggests that this interplay affects the quasi-steady configuration of distributions with ELN-XLN crossings in a larger region of our parameter space. Likewise, the cascade of flavor waves towards narrow angular scales is enhanced. However, flavor conversion in ensembles with $d_v=0$ is still suppressed at $\mu = 630$km$^{-1}$.

\subsection{Mass ordering}
{The influence of the neutrino mass ordering on slow conversion has been investigated  in  e.g.~Ref.~\cite{Raffelt:2007yz} for homogeneous systems, and in the presence of advection in Ref.~\cite{Padilla-Gay:2025tko}. Both works consider monochromatic ensembles and show that the growth rate of flavor coherence is generally higher in IO. However, while the anisotropic example of Ref.~\cite{Raffelt:2007yz} reaches equipartition in both hierarchies, because it starts from equal densities of neutrinos and antineutrinos, the overall amount of flavor conversion  is slightly greater in NO in  Ref.~\cite{Padilla-Gay:2025tko}. Here,} we investigate the difference in the quasi-steady flavor configuration achieved in each neutrino mass ordering for the temperatures in Table~\ref{tab:parameters} and the angular distributions in Figs.~\ref{fig:parameter_space} and \ref{fig:app_error}. 

\begin{figure}[t]
    \centering
    \includegraphics[width=\linewidth]{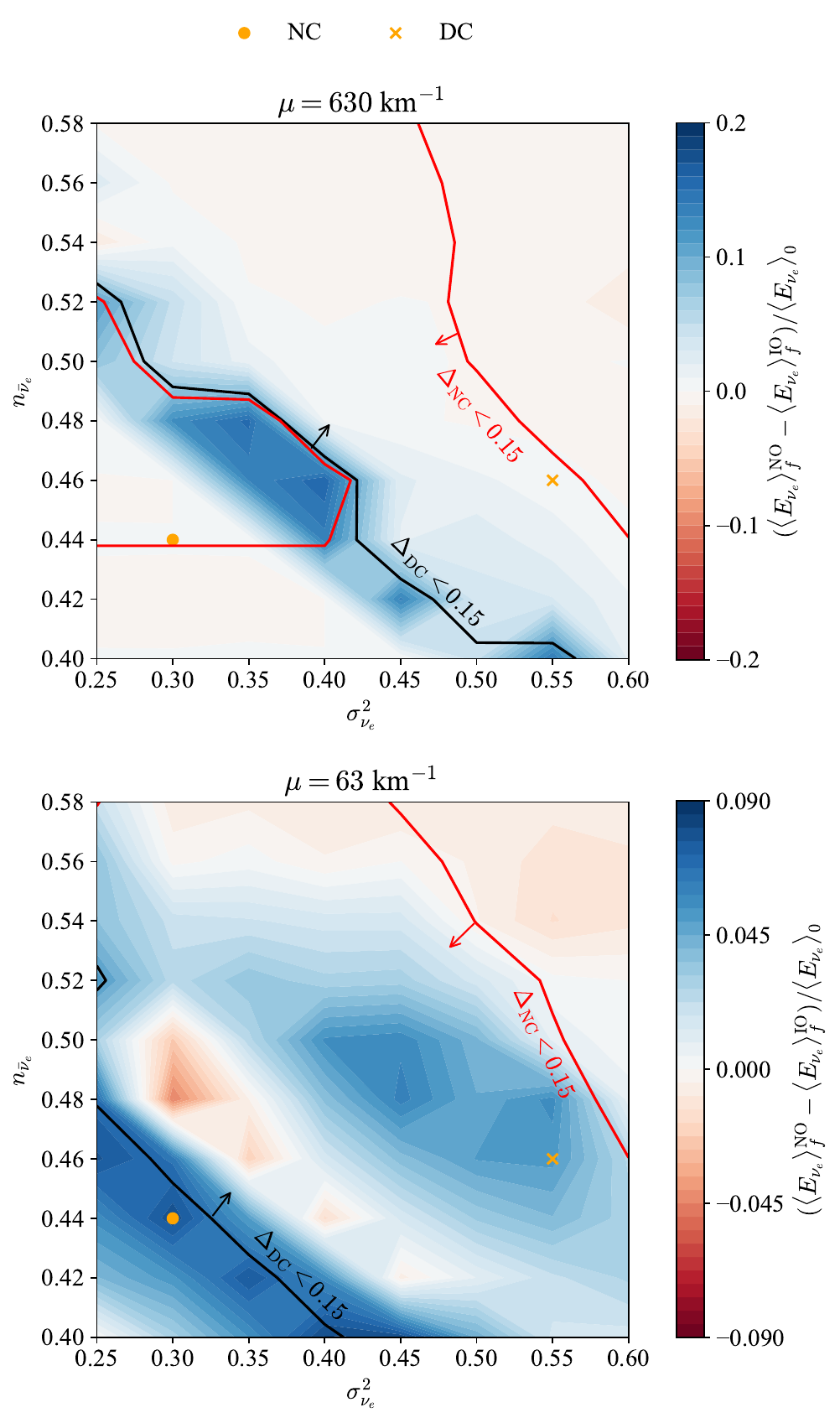}
    \caption{Same as Fig.~\ref{fig:app_error}, but the heat map shows the difference between the  $\nu_e$ average energy after flavor conversion in NO and IO. 
    The difference between the quasi-steady flavor configurations in NO and IO shrinks with increasing $\mu$. The semi-analytic approximation for slow (fast) flavor conversion presented in Sec.~\ref{sec:ansatz_slow} (\ref{sec:ansatz_fast}) agree with the numerical solutions of the kinetic equations in both hierarchies. }
    \label{fig:mass-ordering}
\end{figure}

Figure~\ref{fig:mass-ordering} compares $\langle E_{\nu_e}\rangle_f$ in NO and IO. Overall, ensembles without angular crossings are stable for a broader range of initial parameters in IO. The (anti)neutrino configurations in the lower-left corner of the parameter space remain  frozen in their initial state for $\mu = 63$~km$^{-1}$ (bottom panel):  a $10\%$ decrease compared to the NO implies $\langle E_{\nu_e}\rangle_f^\mathrm{IO}\simeq \langle E_{\nu_e}\rangle_0$. In distributions on the verge of developing an angular crossing, or with  marginal crossings, the amount of flavor conversion in IO is comparable to the NO case or greater, but the evolution is also suppressed more strongly for the higher $\mu$ (top panel), cf.~the diagonal strip extending from ($\sigma^2_{\nu_e}=0.3,\,n_{\bar{\nu}_e}=0.50$) to ($\sigma^2_{\nu_e}=0.55,\, n_{\bar{\nu}_e}=0.40$). In ensembles with deeper ELN-XLN crossings, the IO suppresses flavor conversion by a few percent of $\langle E_{\nu_e}\rangle_0$. The difference with respect to the NO case shrinks with increasing $d_v$ and vanishes in the upper-right corner of our parameter space ($d_v\simeq 1$). The evolution of these ensembles is driven by fast instabilities, which are insensitive to the sign of the vacuum term. Increasing the self-interaction strength brings the ensemble closer to the $\omega \to 0$ limit, in which the solutions of Eqs.~\eqref{eq:qke-vac}  converge for both mass orderings. 

The red (black) contour in Fig.~\ref{fig:mass-ordering} shows the error between our semi-analytic approximation for slow (fast) flavor conversion and the numerical solution in  IO. The quasi-steady states of unstable configurations follow the same qualitative trends as those described in Sec.~\ref{sec:ansatz}. Since the overall amount of flavor conversion is slightly suppressed in IO with respect to the NO case, the cascade of flavor waves towards small angular scales is also less prominent in  IO; this leads to  energy-integrated configurations similar to the Gaussian shape of our ansatz. Hence, the relative error between the semi-analytic expressions and the solution of Eqs.~\eqref{eq:qke-vac} is comparable to that shown in Fig.~\ref{fig:app_error}. 

\section{Discussion and conclusions}
\label{sec:conclusions}
Neutrino-neutrino interactions drive flavor conversion physics in dense environments such as core-collapse supernovae. However, solving quantum neutrino transport numerically exceeds current computational capabilities.
By investigating the quasi-steady flavor configuration due to  slow and fast flavor conversion, we aim to inform subgrid models to be  employed in large-scale simulations of compact astrophysical objects.
We present semi-analytic approximations of the quasi-steady state flavor configuration achieved by 
multi-angle and multi-energy neutrino ensembles in quasi-homogeneous systems. 

We explore the flavor configuration achieved for both mass orderings and  two different values of the self-interaction strength, assuming thermal (anti)neutrino spectra with $e$--$x$ crossings, varying the (anti)neutrino temperatures and the depth of $\nu_e$--$\bar{\nu}_e$ crossings in the energy-integrated ELN-XLN distribution. 

{In the parameter space of (anti)neutrino ensembles that we have explored, the asymptotic amount of flavor conversion tends to be lower in IO.} This finding is consistent with the trend reported in Ref.~\cite{Padilla-Gay:2025tko} for monochromatic neutrinos, but we observe a stronger suppression of flavor conversion for similar ratios of the self-interaction strength to the typical vacuum frequency. 
{
Our results show that the impact of the mass ordering is influenced by the self-interaction strength and the shape of the energy-integrated distributions: if the overall (anti)neutrino density is sufficiently low, flavor conversion can be substantially stronger in IO for ELN-XLN configurations with a shallow angular crossing, or on the verge of one.}

{We find that equipartition at high energies is a good approximation for slow flavor conversion; 
our results may be representative of a finite region of the parameter space in which the $\bar{\nu}_e$ and $\bar{\nu}_x$ spectra are sufficiently close. We provide a generalization of this ansatz to an arbitrary antineutrino survival probability in Eqs.~\eqref{eq:approximation_fast_eb} and \eqref{eq:approximation_fast_xx} for the case of fast conversion: the same functional form could be applied to ensembles without angular crossings in cases with larger $\bar{\nu}_e$--$\bar{\nu}_x$ asymmetries}. These outcomes are qualitatively distinct from the bulb model, where flavor conversion swaps the $\nu_e$ and $\nu_x$ ($\bar{\nu}_e$ and $\bar{\nu}_x$) spectra {within the constraints of lepton-number conservation}~\cite{Hannestad:2006nj,Duan:2006an, Duan:2007bt, Fogli:2007bk}.
The bulb model describes steady neutrino emission from a spherically symmetric object; 
{the stability conditions and flavor evolution of such a system are intrinsically different from those of a quasi-homogeneous box.} 

The errors between our ansatz of the quasi-steady flavor configurations and the numerical solution of the neutrino kinetic equations are $\mathcal{O}(10\%)$  for both neutrino mass orderings. Our semi-analytic approximation of the quasi-steady   configuration due to fast flavor conversion is  qualitatively different from alternative prescriptions, which either assume instantaneous equipartition or impose that fast instabilities  erase angular crossings--see, e.g., Refs.~\cite{Zaizen:2022cik,Xiong:2023vcm,Akaho:2025giw,Richers:2024zit,Wang:2025nii}.  In Ref.~\cite{Goimil-Garcia:2025ozm}, we stressed that flavor equipartition, whether in the entire angular distribution or in a certain angular range, is not a generic outcome of fast instabilities. In this work, we reach similar conclusions for multi-energy (anti)neutrino ensembles. 

{Finally, we discuss some limitations of the prescriptions in this work.} Our assumption that the neutrino survival probabilities are isotropic also makes the piecewise approximations in Eqs.~\eqref{eq:approximation_slow_ee} and \eqref{eq:approximation_fast_ee} discontinuous at high energies. Although  the errors cancel out after integrating over $v$ and are small for thermal spectra, it is worth exploring potential improvements  to enhance the predictability of the quasi-steady state for  codes with discrete energy grids. 
Moreover, we have limited our investigation to flavor configurations with a positive electron lepton number and zero chemical potential. 
More work is needed to extend our semi-analytic ansatz to the case of  binary neutron star merger remnants, where the density of $\bar{\nu}_e$'s can exceed that of $\nu_e$'s. {Last, we assume that the initial (anti)neutrino configurations are separable functions of energy and angle. 
However, the width of the angular distributions should decrease with energy, following the scattering rates: see e.g. Ref.~\cite{Tamborra:2017ubu}.  The ansatz for slow conversion in Eqs.~\eqref{eq:approximation_slow_ee}-\eqref{eq:approximation_slow_xx} only depends on angle-integrated quantities, so the procedure in Sec.~\ref{sec:ansatz_slow} can still be applied in  cases where the (anti)neutrino distributions display correlations in angle and energy. The approximation for fast conversion in Sec.~\ref{sec:ansatz_fast} rescales the antineutrino distributions by constant factors, hence Eqs.~\eqref{eq:approximation_fast_eb} and \eqref{eq:approximation_fast_xx} also hold.  The prescriptions for $\nu_e$ and $\nu_x$ may need to  be generalized for non-trivial scenarios. For example, if the initial neutrino configurations are of the form $\rho_{v,0,\alpha\alpha}(E)\propto f_\alpha(E,T_{\nu_\alpha})G_\alpha(E,v)$, where $\int \mathrm{d}v\,G_\alpha (E,v)=1$, then the first term in Eq.~\eqref{eq:fast_eqv1} should be corrected via $G_\alpha(v=1) \to \int \mathrm{d}E\,F(E,\mathcal{T})G_\alpha (E,v=1)\Theta (E_\mathrm{th}-E)$. The accuracy of the prescriptions would depend on the functional form of $G_\alpha(E,v)$. Optimizing them for different initial conditions beyond those analyzed in this work would require dedicated work. However,  the results presented in this paper demonstrate that the general strategy is robust, matching the numerical solution of the quantum kinetic equation for a broad parameter space.}

\acknowledgments 
This project has received support from the European Union (ERC, ANET, Project No.~101087058).
Views and opinions expressed are those of the authors only and do not necessarily reflect those of the European Union or the European Research Council. Neither the European Union nor the granting authority can be held responsible for them. The Tycho supercomputer hosted at the SCIENCE HPC center at the University of Copenhagen was used to perform the numerical simulations whose results are presented in this paper.

\appendix
\begin{figure*}[t]
    \centering
    \includegraphics[scale=0.8]{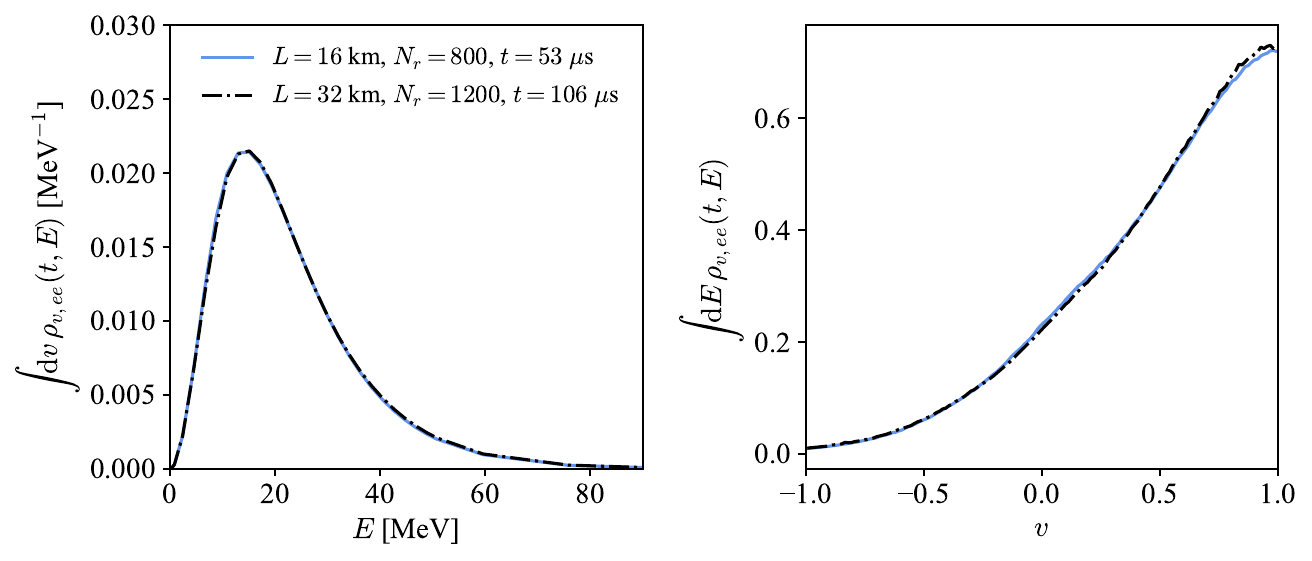}
    \caption{Influence of the size of the periodic box ($L$) and the number of spatial bins ($N_r$) on the quasi-steady state of ensemble DC. \textit{Left:} angle-integrated energy spectrum of $\nu_e$. \textit{Right:} energy-integrated angular distribution of $\nu_e$. Increasing the resolution  has a negligible impact on the space-averaged flavor configuration, showing that the simulation has converged.}
    \label{fig:error_appendix}
\end{figure*}

\section{Radial resolution}
\label{sec:appendix_error}
{In momentum space, the self-interaction term in the equations of motion turns into a convolution of all Fourier modes of the (anti)neutrino density matrices, $\varrho_{v}(t,E,r)$ and $\bar{\varrho}_{v}(t,E,r)$. Consequently, flavor conversion cascades towards smaller spatial scales over time, see e.g.~Ref.~\cite{Goimil-Garcia:2025ozm}. Here, we briefly demonstrate that the resolution adopted in this work is enough to calculate the space-averaged (anti)neutrino distributions, i.e., the homogeneous  mode of $\varrho_{v,\alpha\alpha}$ and $\bar{\varrho}_{v,\alpha\alpha}$.

Figure~\ref{fig:error_appendix} shows the space-averaged energy spectrum (left) and angular distribution (right) of $\nu_e$ for the initial ensemble DC at the end of two different simulations: the solid blue line reproduces the results in Fig.~\ref{fig:approximation_fast}, obtained with a bin size $\Delta r \sim 20\,\mathrm{m}$; the dash-dotted black line shows the numerical solution of Eqs.~\eqref{eq:qke-vac} at a later time, in a bigger box with $33\%$ lower resolution. The results match almost exactly, indicating that the quasi-steady state is independent of the absolute size of the simulation domain ($L$) and that the number of spatial bins adopted in this work is sufficient to resolve flavor conversion at the relevant scales. Comparing the solutions at $t=L$ in boxes with different length also proves that the system has  reached a quasi-steady state.}

\bibliography{Paper}
\end{document}